\documentclass[12pt]{article}
\usepackage{amssymb,amsmath,epsfig}
\usepackage{graphicx}
\allowdisplaybreaks

\begin{document}

\title{\bf Isotropization and Complexity of Decoupled Solutions
in Self-interacting Brans-Dicke Gravity}

\author{M. Sharif \thanks{msharif.math@pu.edu.pk} and Amal Majid
\thanks{amalmajid89@gmail.com}\\
Department of Mathematics, University of the Punjab,\\
Quaid-e-Azam Campus, Lahore-54590, Pakistan.}

\date{}

\maketitle
\begin{abstract}
The aim of this work is to formulate two new solutions by decoupling
the field equations via a minimal geometric deformation in the
context of self-interacting Brans-Dicke gravity. We introduce an
extra source in the anisotropic fluid distribution to generate new
analogs of existing solutions. The radial metric function is
transformed to decouple the field equations into two sets such that
each array corresponds to one source only. The system corresponding
to the original matter distribution is specified by metric functions
of well-behaved solutions. On the other hand, the second set is
closed by imposing constraints on the additional matter source. For
this purpose, we have applied the isotropization condition as well
as vanishing complexity condition on the new source. Smooth matching
of interior and exterior spacetimes at the junction provides values
of the unknown constants. Interesting physical features of
corresponding models are checked by employing the mass and radius of
the star PSR J1614-2230. It is concluded that both extensions yield
viable and stable models for certain values of the decoupling
parameter.
\end{abstract}

\section{Introduction}

The theory of general relativity (GR) has played a significant role
in understanding the mechanism and structure of the vast universe.
The action of a theory encompasses the necessary elements describing
a physical system and its motion. The addition of an extra degree of
freedom in the form of a scalar, vector or tensor field modifies the
action and yields alternative theories of gravity. In 1872, Mach
presented a principle that referred to a relation between the
inertia of a body and matter content in the universe. Later, Dirac's
study of different physical constants alongside cosmological
quantities indicated that the gravitational constant $(G)$ has an
underlying relationship with cosmic time \cite{1}. Their results
motivated Brans and Dicke \cite{2} to modify GR by introducing a
massless scalar field ($\sigma$) incorporating the effects of
dynamic gravitational constant. This scalar-tensor theory (known as
Brans-Dicke (BD) gravity) couples matter and scalar field through a
coupling parameter ($\omega_{BD}$) which may be assigned different
values to describe various cosmic and astrophysical phenomena.

In cosmology, the inflationary model was proposed to overcome the
flatness and horizon problems in the early cosmos. However,
according to this model, the rapidly expanding universe fails to
transition from the period of inflation into the next cosmological
era. Although BD gravity has significant implications in the field
of cosmology, it fails to successfully solve this problem (termed as
the graceful exit problem). The BD coupling parameter must be less
than 25 to adequately describe the inflationary era which disagrees
with the statistical data. Moreover, in BD theory, the effect of
scalar field decreases for higher values of $\omega_{BD}$.
Consequently, large values of the coupling parameter corresponding
to weak field scenario \cite{3} contradict the small values
admissible in the era of cosmic inflation \cite{4}. For the purpose
of rectifying this discrepancy, a potential function ($V(\varrho)$)
is introduced along with a massive scalar field $(\varrho)$ leading
to self-interacting BD (SBD) theory \cite{5}. After this
modification in BD gravity, all values of $\omega_{BD}$ greater than
$\frac{3}{2}$ are permitted if the mass of the scalar field
($m_\varrho$) is more than $2\times10^{-25}GeV$ \cite{6}.

Researchers have employed various techniques to devise solutions
describing different astrophysical phenomena under the influence of
scalar field. Buchdahl \cite{7} proposed that a family of vacuum
solutions can be generated in BD gravity corresponding to static
axially or spherically symmetric GR configurations. Bruckman and
Kazes \cite{8} investigated extremely dense fluid distributions
described by a linear equation of state. They also showed that any
GR solution with a proportional relationship between pressure and
density can be used to construct an interior solution of BD field
equations \cite{9}. Singh and Rai \cite{10} developed a scheme to
obtain axially symmetric BD structure in presence of charge. Krori
and Bhattacharjee \cite{11} transformed the coordinate system to
formulate a Demia\'{n}ski-type metric. The effect of scalar field on
the rotation and other features of slowly as well as rapidly
rotating neutron stars has been extensively discussed \cite{12, 13}.
Viable and stable anisotropic compact models corresponding to the
MIT bag model equation of state have been computed in the presence
of a massive scalar field \cite{14}. Decoupled solutions via MGD
have also been obtained in the background of SBD and other
modifications of GR \cite{15}.

The vast network of interconnected galactic clusters and stars
contains clues regarding the history of our existence and ultimately
the universe. By analyzing the age and evolution of gaseous stars
and their compact remnants, astrophysicists have gained important
information that can prove beneficial in revealing the mysterious
nature of the cosmic mechanism. For this purpose, researchers have
given considerable attention to the mathematical modeling of stellar
structures by computing solutions of the field equations. The field
equations provide the necessary connection between matter and
geometry of large-scale structures but their non-linearity
complicates the extraction of well-behaved solutions representing
realistic physical structures. Schwarzschild \cite{16} resolved this
problem by considering a simple spherical object. However, cosmic
structures are complex systems with a large number of state
determinants. Therefore, researchers are still attempting to devise
different methods that produce viable stellar models.

In this regard, Ovalle \cite{17} extended the well-known Tolman IV
solution by adding a new source. According to Ovalle's technique,
the system of field equations incorporating the effects of seed and
additional sources are decoupled into two arrays via a minimal
geometric deformation (MGD) in the radial coordinate. The first
array includes the energy-momentum tensor corresponding to the
original fluid distribution only while the influence of extra source
is exclusively described by the second system. The known solution
specifies the first set (leading to reduced degrees of freedom)
while the second system is closed by applying a constraint on the
additional matter source. A linear combination of the solutions
corresponding to each system provides a new solution of the entire
system. According to this approach, the coexisting sources do not
exchange energy and are individually conserved. The scheme of
decoupling via MGD has been extended by introducing linear
deformations in radial as well as temporal metric components
\cite{18}.

The decoupling technique via MGD has facilitated the extraction of
many astrophysical and cosmological solutions. Ovalle \cite{19}
first applied MGD approach to accommodate the extra spatial
dimension in the Randall-Sundrum braneworld. Later, he also used
this technique to compute an analog of Tolman IV solution in
braneworld \cite{20}. Casadio et al. \cite{21} introduced an
extension of MGD in context of Randall-Sundrum braneworld to obtain
a modified Schwarzschild geometry. This technique has been generally
employed to construct anisotropic versions of different isotropic
solutions (Tolman IV \cite{22}, Durgapal-Fuloria \cite{23},
Heintzmann \cite{24}, Krori-Barua \cite{25}) in general relativity
(GR). Panotopoulos et al. \cite{26} applied the technique of MGD to
a cloud of strings. Gabbanelli et al. \cite{27} reconstructed the
Schwarzschild solution with cosmological constant via decoupling and
obtained an extension with non-uniform density. This approach has
been successfully applied in (2+1)-dimensional spacetime as well
\cite{28, 29}.

Casadio et al. \cite{30} used decoupling to control the anisotropy
and complexity of a static self-gravitating system. Hensh and
Stuchlik \cite{31} deformed the radial metric function of Tolman VII
spacetime to evaluate its anisotropic version. Singh et al.
\cite{32} developed isotropic solutions using Karmarkar's condition
and then extended them to the anisotropic domain via MGD. Viable
anisotropic extensions of a cloud of strings have also been obtained
by employing the same technique \cite{33}. Cede\~{n}o and Contreras
\cite{34} showed that the decoupling approach is also applicable in
the field of cosmology by deforming FLRW and Kantowski-Sachs
spacetimes. Tello-Ortiz et al. \cite{35} extended the Durgapal IV
solution to anisotropic domain via MGD. Deformations of Dirac as
well as Yang-Mills-Dirac configurations have also been obtained by
deforming the radial metric function \cite{36, 37}. Axially
symmetric Kerr and Kerr-Newman spacetimes were also successfully
extended via this scheme \cite{38}. Recently, Contreras and
Fuenmayor \cite{39} analyzed the stability of static spheres via
gravitational cracking in the background of decoupling. Sultana
\cite{40} showed the applicability of decoupling through MGD in
higher order theories as well.

In this paper, we show that the decoupling via the MGD scheme can be
adopted to control certain physical features of static spherical
self-gravitating systems in the context of SBD gravity. We compute
two new solutions by governing the anisotropy as well as complexity
of the matter distribution. The paper is arranged in the following
format. The anisotropic field equations involving the additional
source are constructed in the next section. The radial deformation
is applied to decouple the field equations in section \textbf{3}. In
section \textbf{4}, we construct new versions of well-known ansatz
and inspect them graphically. A summary of the main results is
presented in the last section.

\section{Self-interacting Brans-Dicke Theory}

Replacing the gravitational constant in GR by a massive scalar field
leads to the action (in relativistic units) of SBD gravity given as
\begin{equation}\label{0}
\mathcal{S}=\int\sqrt{-g}(\mathfrak{R}\varrho-\frac{\omega_{BD}}
{\varrho}\nabla^{\gamma}\varrho\nabla_{\gamma}\varrho
-V(\varrho)+\emph{L}_m+\eta\emph{L}_\Theta) d^{4}x,
\end{equation}
where the Lagrangians of matter and additional sources are denoted
by $\emph{L}_m$ and $\emph{L}_\Theta$, respectively. The influence
of the extra source ($\Theta$) is determined by the decoupling
parameter $\eta$. Moreover, $g$ and $\mathfrak{R}$ denote the
determinant of the metric tensor ($g_{\gamma\delta}$) and Ricci
scalar, respectively. The field and wave equations in SBD gravity,
respectively read
\begin{eqnarray}\label{1}
G_{\gamma\delta}&=&\mathfrak{R}_{\gamma\delta}-\frac{1}{2}g_{\gamma\delta}\mathfrak{R}
=T^{\text{(eff)}}_{\gamma\delta}=\frac{1}{\varrho}(T_{\gamma\delta}^{(m)}
+\eta\Theta_{\gamma\delta}+T_{\gamma\delta}^\varrho),\\\label{2}
\Box\varrho&=&\frac{1}{2\omega_{BD}+3}\left(T^{(m)}+\eta\Theta+
(\varrho\frac{dV(\varrho)}{d\varrho}-2V(\varrho))\right).
\end{eqnarray}
Here, $T_{\gamma\delta}^{(m)}$ corresponds to the energy-momentum
tensor of the original fluid distribution while the impact of the
scalar field on salient characteristics of the compact structure is
measured through the tensor
\begin{equation}\label{4}
T_{\gamma\delta}^\varrho=\varrho_{,\gamma;\delta}-g_{\gamma\delta}
\Box\varrho+\frac{\omega_{BD}}{\varrho}
(\varrho_{,\gamma}\varrho_{,\delta}
-\frac{g_{\gamma\delta}\varrho_{,\alpha}\varrho^{,\alpha}}{2})
-\frac{V(\varrho)g_{\gamma\delta}}{2},
\end{equation}
where $\Box$ is the d'Alembertian operator,
$g^{\gamma\delta}T_{\gamma\delta}^{(m)}=T^{(m)}$ and
$g^{\gamma\delta}\Theta_{\gamma\delta}^{(m)}=\Theta$. The two matter
sources are present in a static spherical region covered by a
three-dimensional hypersurface $(\Sigma)$ and described by the line
element
\begin{equation}\label{5}
ds^2=e^{\phi_1(r)}dt^2-e^{\phi_2(r)}dr^2-r^2d\theta^2-r^2\sin^2\theta
d\varphi^2.
\end{equation}

We assume that the seed source is anisotropic in nature whose energy
and momentum are represented by the following tensor
\begin{equation}\label{3}
T_{\gamma\delta}^{(m)}=(\rho+p_\perp)
v_{\gamma}v_{\delta}-p_{\perp}g_{\gamma\delta}+(p_r-p_\perp)s_\gamma
s_\delta,
\end{equation}
where $\rho$, $p_r$ and $p_\perp$ denote the energy density, radial
and transverse components of pressure, respectively. Moreover,
$v_\gamma=(e^{\frac{\phi_1}{2}},~0,~0,~0)$ is the four-velocity
while $s_\gamma=(0,~-e^{\frac{\phi_2}{2}},~0,~0)$ corresponds to the
radial four-vector. The effective energy-momentum tensor,
incorporating the effects of scalar field and additional source,
obeys the conservation law. The field equations (\ref{1}) take the
following form for metric (\ref{5}) in the presence of scalar field
\begin{eqnarray}\label{6}
\frac{1}{r^2}-e^{-\phi_2}\left(\frac{1}{r^2}-\frac{\phi_2'}{r}\right)
-\frac{T_0^{0\varrho}}{\varrho}&=&
\frac{1}{\varrho}(\rho+\eta\Theta_0^0),\\\label{7}
-\frac{1}{r^2}+e^{-\phi_2}\left(\frac{1}{r^2}+\frac{\phi_1'}{r}\right)
+\frac{T_1^{1\varrho}}{\varrho}&=&
\frac{1}{\varrho}(p_r-\eta\Theta_1^1),\\\label{8}
\frac{e^{-\phi_2}}{4}\left(2\phi_1''+\phi_1'^2-\phi_2'\phi_1'+2\frac{\phi_1'-\phi_2'}{r}\right)
+\frac{T_2^{2\varrho}}{\varrho}&=&
\frac{1}{\varrho}(p_\perp-\eta\Theta_2^2),
\end{eqnarray}
where
\begin{eqnarray}\nonumber
T_0^{0\varrho}&=&e^{-\phi_2}\left[\varrho''+\left(\frac{2}{r}-\frac{\phi_2'}{2}
\right)\varrho'+\frac{\omega_{BD}}{2\varrho}\varrho'^2-e^{\phi_2}\frac{V(\varrho)}
{2}\right],\\\nonumber
T_1^{1\varrho}&=&e^{-\phi_2}\left[\left(\frac{2}{r}+\frac{\phi_1'}
{2}\right)\varrho'-\frac{\omega_{BD}}{2\varrho}\varrho'^2
-e^{\phi_2}\frac{V(\varrho)}{2})\right],\\\nonumber
T_2^{2\varrho}&=&e^{-\phi_2}\left[\varrho''+\left(\frac{1}{r}-\frac{\phi_2'}
{2}+\frac{\phi_1'}{2}\right)\varrho'+\frac{\omega_{BD}}{2\varrho}\varrho'^2
-e^{\phi_2}\frac{V(\varrho)}{2} \right].
\end{eqnarray}
Here, differentiation with respect to $r$ is indicated by $'$.
Moreover, the signature is included in the unknown components of
$\Theta_{\delta}^{\gamma}$. Furthermore, the wave equation
corresponding to the current scenario is expressed as
\begin{eqnarray}\nonumber
\Box\varrho&=&-e^{-\phi_2}\left[\left(\frac{2}{r}-\frac{\phi_2'} {2}
+\frac{\phi_1'}{2}\right)\varrho'+\varrho''\right]\\\label{2*}
&=&\frac{1}{3+2\omega_{BD}}\left[\eta\Theta+T^{(m)}+
\left(\varrho\frac{dV(\varrho)}{d\varrho}-2V(\varrho)\right)\right].
\end{eqnarray}

Various cosmological and astrophysical models have been developed
corresponding to different forms of self-interacting potentials
\cite{41}. The specific form of potential is unknown but it has been
assumed that at high temperatures the potential is proportional to
$\varpi^{2}$ \cite{42}. Furthermore, Quiros \cite{43} showed that
the de Sitter solution in general relativity arises in the SBD
theory for the quadratic potential ($V(\varpi)=m^2\varpi^{2}$, where
$m$ is coupled to the mass of the scalar field) only. We proceed by
assuming that
\begin{equation*}
V(\varpi)=\frac{1}{2}m_{\varpi}^2\varpi^2,
\end{equation*}
where $m_\varrho=0.01$ as the weak field observations are satisfied
in SBD theory for $m_{\varrho}> 10^{-4}$ (in dimensionless units).
Note that the combination
$(\varpi\frac{dV(\varpi)}{d\varpi}-2V(\varpi))$ in Eq.(\ref{2*})
vanishes for the chosen potential function which simplifies the
mathematical calculations. This form of potential function has been
previously considered for different cosmic scenarios \cite{13, 42,
44}. It is noted that the inclusion of the source $\Theta$ in the
field equations has increased the number of unknowns:
$\phi_1,~\phi_2,~\rho,~p_r,~p_\perp,~\Theta_0^0,~\Theta_1^1,~\Theta_2^2,~\varrho$.

\section{Gravitational Decoupling}

The scheme of decoupling via MGD transforms the radial metric
component as
\begin{eqnarray}\label{11}
e^{-\phi_2(r)}\mapsto \lambda(r)+\eta \mu(r),
\end{eqnarray}
while the temporal metric potential is not deformed. The
modification in the radial component is controlled by the
deformation function $\mu(r)$. Moreover, the linear mapping does not
disturb the symmetry of the sphere. As a consequence of applying the
transformation, the complicated field equations yield two arrays.
The first set (for $\eta=0$) encodes the influence of the
anisotropic seed source as
\begin{eqnarray}\nonumber
\rho&=&-\frac{1}{2r^2\varrho(r)}[r^2\omega_{BD}\lambda(r)\varrho'^2-r^2\varrho(r)V(\varrho)
+r\varrho(r)(r\lambda'(r)\varrho'(r)+2r\lambda\varrho''\\\label{12}
&+&4\lambda(r)\varrho'(r))+2\varrho^2(r)
(r\lambda'(r)+\lambda(r)-1)],\\\nonumber
p_r&=&\frac{1}{r^2}[\varrho(r)(r\lambda(r)
\phi_1'(r)+\lambda(r)-1)]+\frac{1}{2r\varrho(r)}[\lambda(r)\varrho'(r)(\varrho(r)(r\phi_1'(r)+4)\\\label{13}
&-&r\omega_{BD}\varrho'(r))]-\frac{V(\varrho)}{2},\\\nonumber
p_\perp&=&\frac{1}{4r\varrho(r)}[\varrho(r)\lambda'(r)(\varrho(r)(r\phi_1'(r)+2)
+2r\varrho'(r))+\lambda(r)(2\varrho(r)\varrho'(r)\\\nonumber
&\times&((r\phi_1'(r)+2)+2r\varrho''(r))+\varrho^2(r)(2r\phi_1''(r)+r\phi_1'^2(r)+2\phi_1'(r))\\\label{14}
&+&2r\omega_{BD}\varrho'^2(r))-2r\varrho(r)V(\varrho)].
\end{eqnarray}
The conservation equation of the seed matter distribution is
obtained as
\begin{equation}\label{14*}
T^{1'(\text{eff})}_{1}-\frac{\phi_1'(r)}{2}
(T^{0(\text{eff})}_{0}-T^{1(\text{eff})}_{1})
-\frac{2}{r}(T^{2(\text{eff})}_{2}-T^{1(\text{eff})}_{1})=0.
\end{equation}
The characteristics of the additional source are exclusively
included in the second set as
\begin{eqnarray}\nonumber
\Theta_0^0&=&\frac{-1}{2r^2\varrho(r)}[r\varrho(r)\mu'(r)(r\varrho'(r)+2\varrho(r))+\mu(r)
(r^2\omega_{BD}\varrho'^2(r)+2r\varrho(r)\\\label{15}
&\times&(r\varrho''(r)+2\varrho'(r))+2\varrho^2(r))],\\\nonumber
\Theta_1^1&=&\frac{-1}{2r^2\varrho(r)}[\mu(r)(-r^2\omega_{BD}\varrho'^2(r)+r\varrho(r)(r
\phi_1'(r)+4)\varrho'(r)+2\varrho^2(r)\\\label{16}
&\times&(r\phi_1'(r)+1))],\\\nonumber
\Theta_2^2&=&\frac{-1}{4\varrho(r)}[2\varrho(r)(r
\mu'(r)\varrho'(r)+\mu(r)((r\phi_1'(r)+2)\varrho'(r)+2r\varrho''(r)))\\\nonumber
&+&\varrho^2(r)(\mu'(r)(r\phi_1'(r)+2)+\mu(r)(2r\phi_1''(r)+r\phi_1'^2(r)+2\phi_1'(r)))\\\label{17}
&+&2r\omega_{BD}\mu(r)\varrho'^2(r)].
\end{eqnarray}

The conservation equation associated with the above system is
\begin{equation}\label{17*}
\Theta^{1'(\text{eff})}_{1}-\frac{\phi_1'(r)}{2}(\Theta^{0(\text{eff})}_{0}
-\Theta^{1(\text{eff})}_{1})
-\frac{2}{r}(\Theta^{2(\text{eff})}_{2}-\Theta^{1(\text{eff})}_{1})=0,
\end{equation}
where
\begin{eqnarray*}
\Theta^{0(\text{eff})}_0&=&\frac{1}{\varrho}\left(\Theta^0_0+\frac{1}{2}
\mu'(r)\varrho'(r)+\mu(r)\varrho''(r)+\frac{\omega_{BD}\mu(r)\varrho'^2(r)}{2\varrho(r)}
+\frac{2\mu(r)\varrho '}{r}\right),\\
\Theta^{1(\text{eff})}_1&=&\frac{1}{\varrho}\left(\Theta^1_1+\frac{1}{2}
\mu(r)\phi_1'(r)\varrho'(r)-\frac{\omega_{BD}\mu(r)\varrho'^2(r)}{2\varrho(r)}+\frac{2
\mu(r)\varrho'(r)}{r}\right),\\
\Theta^{2(\text{eff})}_2&=&\frac{1}{\varrho}\left(\Theta^2_2+\frac{1}{2}
\mu'(r)\varrho'(r)+\frac{1}{2}\mu(r)\phi_1'(r)\varrho'(r)+\mu(r)\varrho''(r)\right.\\
&+&\left.\frac{\omega_{BD}\mu(r)\varrho'^2(r)}{2\varrho(r)}+\frac{\mu(r)\varrho'(r)}{r}\right).
\end{eqnarray*}
The MGD technique prohibits the exchange of energy between the
additional and original fluid distributions. Thus, the two sources
are conserved individually as indicated by Eqs.(\ref{14*}) and
(\ref{17*}). The number of unknown variables decreases if a
well-behaved solution is assumed for Eqs.(\ref{12})-(\ref{14}). On
the other hand, a constraint on $\Theta$-sector determines the
deformation function and specifies the second set. Finally,
combining these solutions generates new ansatz with the following
energy density and pressure components
\begin{eqnarray}\nonumber
\rho&=&\frac{-1}{2r^2\varrho(r)}[r\varrho(r)(r\varrho'(r)(\eta
\mu'(r)+\lambda'(r))+2\eta
\mu(r)(r\varrho''(r)+2\varrho'(r))\\\nonumber
&+&2\lambda(r)(r\varrho''(r)+2\varrho'(r)))+2\varrho^2(r)(\eta
r\mu'(r)+\eta \mu(r)+r\lambda'(r)+\lambda(r)\\\label{28}
&-&1)+r^2\omega_{BD}\varrho'^2(r)(\eta
\mu(r)+\lambda(r))-r^2\varrho(r)V(\varrho)],\\\nonumber
p_r&=&\frac{1}{2r^2\varrho}[r^2\omega_{BD}\varrho'^2(r)(\eta
\mu(r)-\lambda(r))-r\varrho(r)(r\phi_1'(r)+4)\varrho'(r)(\eta
\mu(r)\\\nonumber &-&\lambda(r))+2\varrho^2(r)(-\mu(r)(\eta +\eta
r\phi_1'(r))+r\lambda(r)\phi_1'(r)+\lambda(r)-1)\\\label{29}
&-&r^2\varrho(r)V(\varrho)],\\\nonumber
p_\perp&=&\frac{1}{4r\varrho(r)}[-2\varrho(r)(r \varrho'(r)(\eta
\mu'(r)-\lambda'(r))+\eta
\mu(r)((r\phi_1'(r)+2)\varrho'(r)\\\nonumber
&+&2r\varrho'')-\lambda(r)((r\phi_1'(r)+2)\varrho'(r)+2r\varrho''(r)))-\varrho^2(r)(\eta
\mu'(r)(r\phi_1'(r)\\\nonumber &+&2)-\eta \mu(2r\phi_1''(r)+r
\phi_1'^2(r)+2
\phi_1'(r))+2r\lambda(r)\phi_1''(r)+r\phi_1'(r)\lambda'(r)\\\nonumber
&+&r\lambda(r)\phi_1'^2(r)+2\lambda(r)\phi_1'(r)+2\lambda'(r))+2r\omega_{BD}\varrho'^2(r)(\lambda(r)-\eta
\mu(r))\\\label{30} &-&2r\varrho(r)V(\varrho )].
\end{eqnarray}

\section{Decoupled Solutions}

In this section, we impose constraints on the new source to develop
two solutions representing self-gravitating systems in the presence
of massive scalar field. We graphically analyze the generated
solutions and inspect their important features.

\subsection{Solution I}

For the first solution, we employ the following ansatz to determine
the set related to anisotropic seed source
\begin{eqnarray}\label{40}
\phi_1(r)&=&\ln\left(B^2
\left(\frac{r^2}{A^2}+1\right)\right),\\\label{41}
\lambda(r)&=&\left(\frac{A^2+r^2}{A^2+3 r^2}\right),
\end{eqnarray}
where $A$ and $B$ are constants. These metric potentials belong to a
class of solutions that determine the gravitational field in a
cluster of particles following randomly oriented circular orbits
\cite{45}. Moreover, this spacetime has been previously used to
construct decoupled solutions in GR \cite{30}. The surface of the
compact object is the junction between the interior and exterior
spacetimes. In order to avoid any singularity, the anisotropic
interior must smoothly connect with the vacuum Schwarzschild metric
defined as
\begin{equation}\label{20*}
ds^2=\frac{r-2M}{r}dt^2-\frac{r}{r-2M}dr^2
-r^2d\theta^2-r^2\sin^2\theta d\varphi^2,
\end{equation}
where $M$ denotes the mass. The form of scalar field in the vacuum
exterior has been determined via the procedure used in \cite{8}. The
conditions
\begin{eqnarray*}
(g_{\gamma\delta}^-)_{\Sigma}&=&(g_{\gamma\delta}^+)_{\Sigma},\quad
(p_{r})_{\Sigma}=0,\\
(\varrho^-(r))_\Sigma&=&(\varrho^+(r))_\Sigma,\quad
(\varrho'^-(r))_\Sigma=(\varrho'^+(r))_\Sigma,
\end{eqnarray*}
ensure the smooth matching at the hypersurface ($\Sigma:r=R$, $R$ is
the radius of the celestial object). The values of unknown constants
are obtained through these conditions as
\begin{eqnarray}\label{20}
A&=&\frac{\sqrt{R^2(R-3M)}}{\sqrt{M}},\\\label{21}
B&=&\frac{\sqrt{R-2M}}{\sqrt{\frac{MR}{R-3M}+R}},
\end{eqnarray}
where $M$ and $R$ are specified by employing the mass and radius of
PSR J1614-2230 ($M=1.97M_{\bigodot}$ and $R=11.29km$). The field
equations (\ref{28})-(\ref{30}) take the following form
corresponding to Eqs.(\ref{40}) and (\ref{41})
\begin{eqnarray}\nonumber
\rho&=&-\left(2 r^2 \left(A^2+3r^2\right)^2
\varrho\right)^{-1}\left(2 \varrho^2 \left(r \left(\eta \left(A^2+3
r^2\right)^2 \mu'-6 r
\left(A^2+r^2\right)\right)\right.\right.\\\nonumber
&+&\left.\left.\eta \left(A^2+3r^2\right)^2 \mu(r)\right)+r^2
\omega_{BD} \left(A^2+3 r^2\right) \varrho '^2 (r)\left(\eta
\left(A^2+3 r^2\right) \mu(r)\right.\right.\\\nonumber
&+&\left.\left.A^2+r^2\right)-r^2 \left(A^2+3 r^2\right)^2 \varrho
(r) V(\varrho )+r \varrho (r) \left(2 r \left(A^2+3 r^2\right)
\varrho''(r)\right.\right.\\\nonumber &\times&\left.\left.\left(\eta
\left(A^2+3 r^2\right) \mu(r)+A^2+r^2\right)+\varrho'(r) \left(\eta
r \left(A^2+3 r^2\right)^2 \mu'(r)\right.\right.\right.\\\nonumber
&+&\left.\left.\left.4\eta\left(A^2+3 r^2\right)^2 \mu(r)+4
\left(A^4+3 A^2 r^2+3 r^4\right)\right)\right)\right),\\\nonumber
p_r&=&\frac{-1}{2 r \varrho}\left(\varrho'
\left(\frac{A^2+r^2}{A^2+3 r^2}+\eta \mu\right) \left(\left(-\frac{2
r^2}{A^2+r^2}-4\right) \varrho (r)+r \omega_{BD} \varrho
'\right)\right)\\\nonumber &+&\frac{\eta \left(A^2+3 r^2\right)
\mu(r) \varrho (r)}{r^2 \left(A^2+r^2\right)}-\frac{V(\varrho
)}{2},\\\nonumber p_\perp&=&\left(2 r \left(A^2+r^2\right)^2
\left(A^2+3 r^2\right)^2 \varrho (r)\right)^{-1}\left(\varrho ^2(r)
\left(\left(A^2+r^2\right) \left(\eta \left(A^2+2
r^2\right)\right.\right.\right.\\\nonumber
&\times&\left.\left.\left.\left(A^2+3 r^2\right)^2 \mu'+6
r^3\left(A^2+r^2\right)\right)+2 \eta
r\left(2A^2+r^2\right)\left(A^2+3 r^2\right)^2
\mu(r)\right)\right.\\\nonumber
&+&\left.r\omega_{BD}\left(A^2+r^2\right)^2\left(A^2+3r^2\right)\varrho'^2\left(\eta\left(A^2+3
r^2\right)\mu(r)+A^2+r^2\right)\right.\\\nonumber
&+&\left.\left(A^2+r^2\right) \varrho (r) \left(\varrho '(r) \left(2
\eta  \left(A^2+2 r^2\right) \left(A^2+3r^2\right)^2
\mu(r)+\left(A^2+r^2\right)\right.\right.\right.\\\nonumber
&\times&\left.\left.\left.\left(\eta r \left(A^2+3 r^2\right)^2
\mu'+2 \left(A^4+3 A^2 r^2+6r^4\right)\right)\right)+2 r\varrho ''
\left(A^4+4 A^2 r^2\right.\right.\right.\\\nonumber
&+&\left.\left.\left.3r^4\right)  \left(\eta \left(A^2+3 r^2\right)
\mu+A^2+r^2\right)\right)-r \left(A^4+4 A^2 r^2+3 r^4\right)^2
\varrho V(\varrho )\right).
\end{eqnarray}

The total anisotropy of the system under the influence of scalar
field is
\begin{equation}\label{9}
\Delta=(p_\perp-\eta\Theta_2^2)-(p_r-\eta\Theta_1^1)
=(p_\perp-p_r)+\eta(\Theta_1^1-\Theta_2^2),
\end{equation}
We solve Eqs.(\ref{15})-(\ref{17}) in the current setup by assuming
that the addition of new source converts the anisotropic structure
into an isotropic system for $\eta=1$, i.e., $\Delta$ reduces to
zero for $\eta=1$ yielding
\begin{equation}\label{10}
p_\perp-p_r=-(\Theta_1^1-\Theta_2^2).
\end{equation}
Recently, Casadio et al. \cite{30} utilized the above condition to
isotropize an anisotropic source through decoupling. The
isotropization condition corresponding to the metric functions in
Eqs.(\ref{40}) and (\ref{41}) takes the form
\begin{eqnarray}\nonumber
&&\left(r \varrho\left(A^2+r^2\right) \left(A^2+3 r^2\right)
\right)^{-1}\left(2 r^2 \omega_{BD}\left(A^2+3
r^2\right)\left(A^2+r^2\right)^2\left(\left(A^2\right.\right.\right.\\\nonumber
&&+\left.\left.\left.3r^2\right) \mu+A^2+r^2\right) \varrho '^2+r
\left(A^2+r^2\right)^2 \varrho \left(2 r \left(A^2+3 r^2\right)
\left(\left(A^2+3 r^2\right)\mu\right.\right.\right.\\\nonumber
&&+\left.\left.\left.A^2+r^2\right) \varrho''(r)+\varrho '(r)\left(r
\left(A^2+3 r^2\right)^2 \mu'(r)-2 \left(A^2+3 r^2\right)^2
\mu(r)\right.\right.\right.\\\nonumber
&&-\left.\left.\left.2\left(A^4+6 A^2 r^2+3
r^4\right)\right)\right)+\varrho^2 \left(r \left(A^2+r^2\right)
\left(\left(A^2+2 r^2\right) \left(A^2+3 r^2\right)^2
\mu'\right.\right.\right.\\\label{10*}
&&\left.\left.\left.+6r^3\left(A^2+r^2\right)\right)-2 \left(A^2+3
r^2\right)^2 \left(A^4+2 A^2 r^2+2 r^4\right) \mu\right)\right)=0.
\end{eqnarray}
The deformation function is evaluated by numerically solving the
above equation alongside Eq.(\ref{2*}) for $\varrho(0)=0.1$ and
$\mu(0)=0$. Further, we graphically analyze the constructed solution
for $\eta=0.2,~0.5,~1$. The corresponding values of coupling
parameter are mentioned in Table \textbf{1} where the quantities
with subscripts $c$ and $s$ are calculated at the center and surface
of the star, respectively. Moreover, the constant $A$ does not
change after the transformation while $B$ cannot be determined.
Therefore, we choose the value of $B$ as given in Eq.(\ref{21}). The
transformed radial metric function is monotonically increasing and
free from singularity as shown in Figure \textbf{1}. Compact objects
are the most dense at the center leading to maximum pressure at
$r=0$. Moreover, the radial pressure always approaches to zero near
the boundary of the cosmic structure. Figure \textbf{2} indicates
that the matter variables are consistent with the required criteria.
The object becomes more dense as the value of decoupling parameter
increases from 0.2 to 0.5 but then decreases for $\eta=1$. However,
pressure components follow the opposite trend, i.e.,
radial/tangential pressure decreases when value of $\eta$ increases
from 0.2 to 0.5 but then slightly increases for $\eta=1$. It is
noted that anisotropy is zero at the center for $\eta=0.2,~0.5$
whereas it vanishes for $\eta=1$ as required.
\begin{table} \caption{Values of physical parameters in solution I corresponding to
$\eta=0.2,~0.5,~1$.}
\begin{center}
\begin{tabular}{|c|c|c|c|}
\hline &$\eta=0.2$ & $\eta=0.5$ & $\eta=1$  \\
\hline $\omega_{BD}$ &-1.4 & 6 & 1 \\
\hline $\rho_c~(gm/cm^3)$ & $4.506\times10^{15}$ &
$7.721\times10^{15}$ & $4.385\times10^{15}$\\
\hline$\rho_s~(gm/cm^3)$ & $4.883\times10^{14}$ &
$1.171\times10^{15}$&$2.448\times10^{14}$\\
\hline $p_c~(dyne/cm^2)$ & $1.3\times10^{36}$ &
$4.148\times10^{36}$ & $7.465\times10^{36}$\\
\hline$Z_s$ &0.026 & 0.042 & 0.0218
\\\hline
\end{tabular}
\end{center}
\end{table}
\begin{figure}\center
\epsfig{file=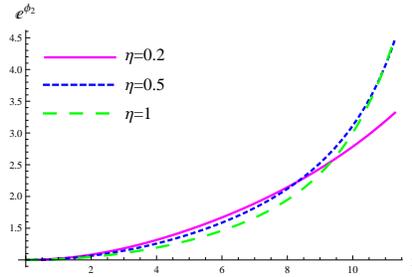,width=0.4\linewidth} \caption{Behavior of
deformed radial metric potential for solution I.}
\end{figure}
\begin{figure}\center
\epsfig{file=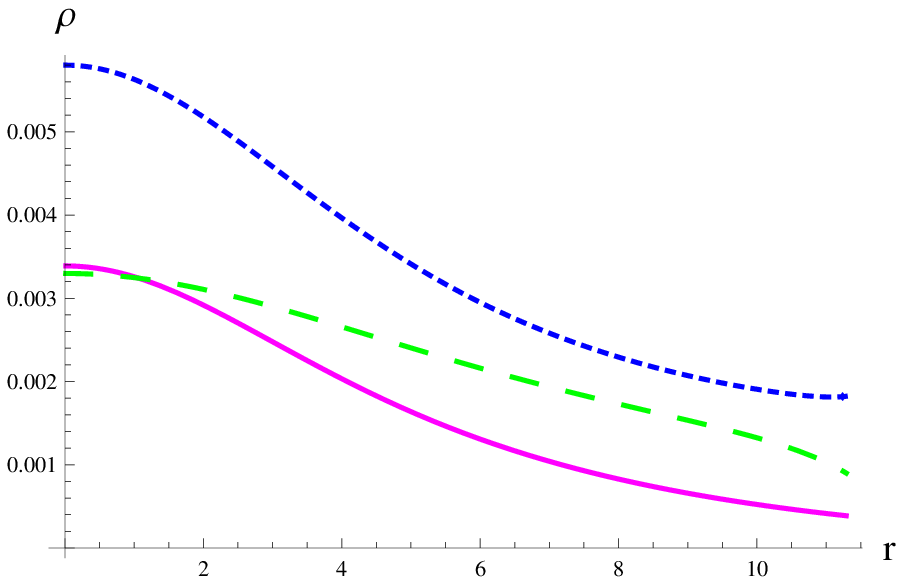,width=0.4\linewidth}\epsfig{file=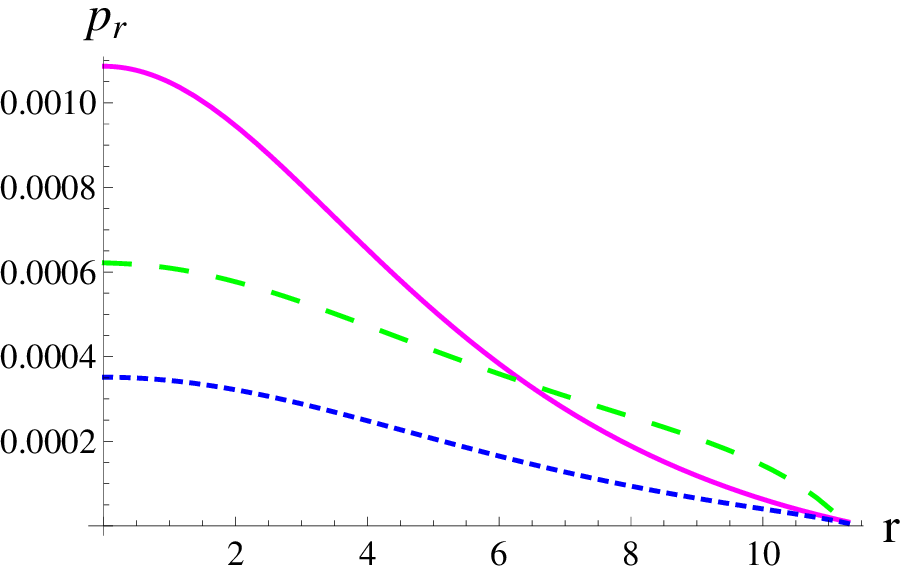,width=0.4\linewidth}
\epsfig{file=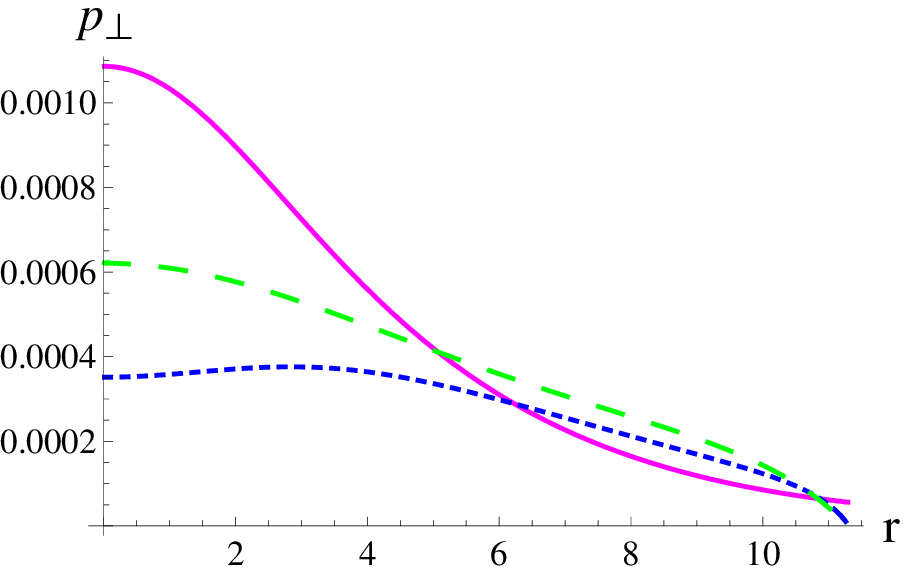,width=0.4\linewidth}\epsfig{file=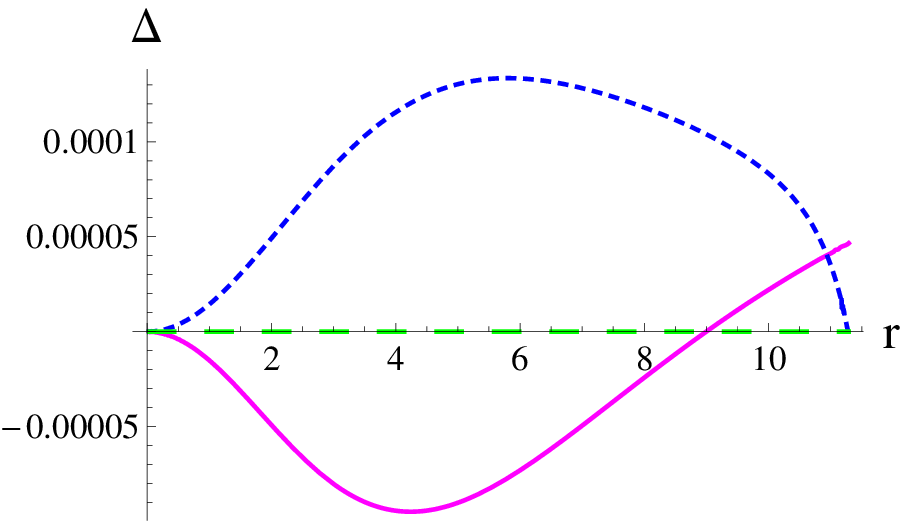,width=0.4\linewidth}
\caption{Behavior of $\rho,~p_r,~p_\perp$ (in $km^{-2}$) and
$\Delta$ for solution I.}
\end{figure}
\begin{figure}\center
\epsfig{file=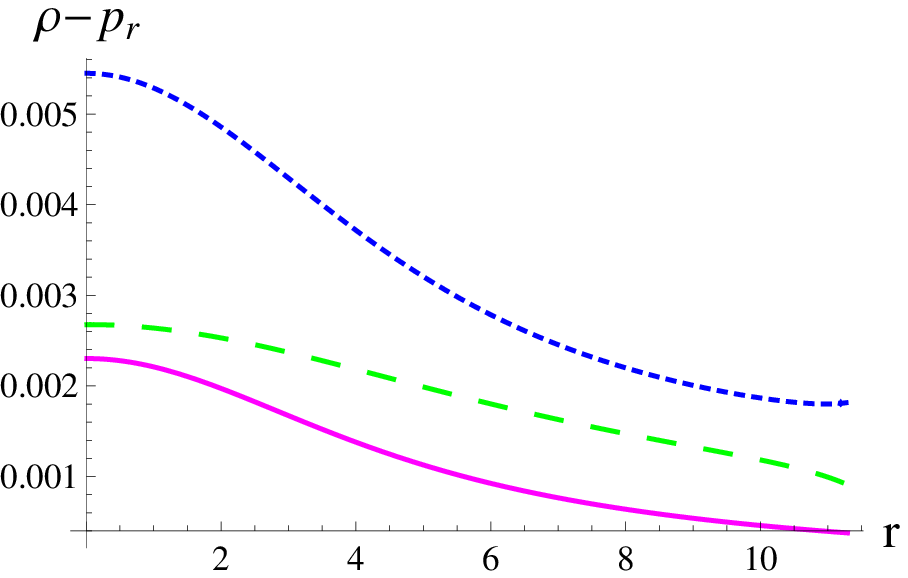,width=0.4\linewidth}\epsfig{file=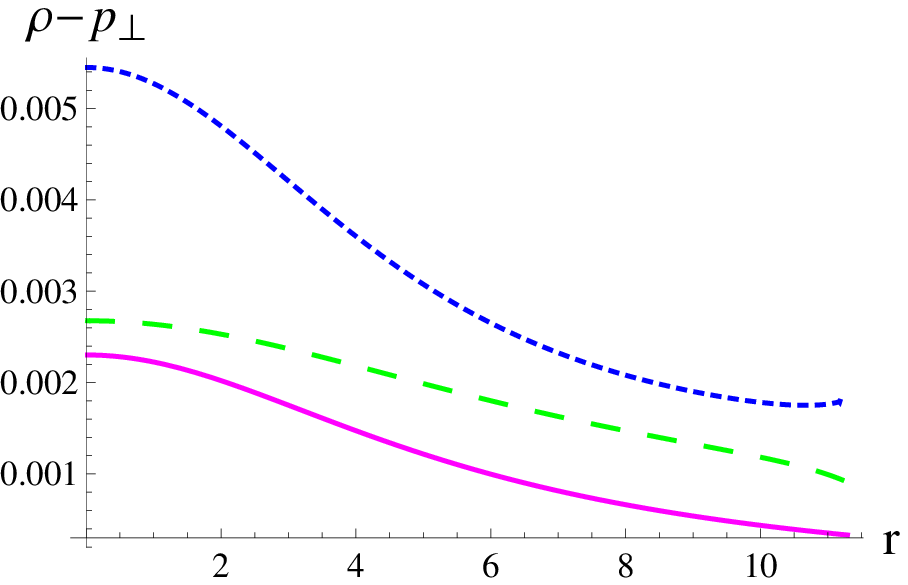,width=0.4\linewidth}
\caption{DEC for solution I.}
\end{figure}

The astrophysical systems composed of normal matter comply with four
energy bounds listed as \cite{46}
\begin{eqnarray*}
&&\text{null energy condition:}\quad\rho+p_r\geq0,\quad\rho+p_\perp\geq0,\\
&&\text{weak energy condition:}\quad\rho\geq0,\quad\rho+p_r\geq0,\quad\rho+p_\perp\geq0,\\
&&\text{strong energy condition:}\quad\rho+p_r+2p_\perp\geq0,\\
&&\text{dominant energy condition:}\quad\rho-p_r\geq0,\quad
\rho-p_\perp\geq0.
\end{eqnarray*}
These conditions ensure that the fluid distribution inside the
static sphere is physically viable. The new structure satisfies the
first three energy bounds as the density and pressure are
non-negative throughout the interior (refer to Figure \textbf{2}).
Further, the graphical representations of $\rho-p_r$ and
$\rho-p_\perp$ in Figure \textbf{3} indicate the fulfilment of
dominant energy condition as well. Thus, the matter distribution
within spherical structure is viable for the considered values of
$\eta$.

In order to compute the mass of the spherical distribution, we
numerically solve the differential equation
\begin{equation}\label{31}
\frac{dm(r)}{dr}=\frac{1}{2}r^2\rho,
\end{equation}
subject to the initial condition $m(0)=0$. The compactness of the
celestial object depends on the arrangement of composite particles.
Highly compact objects have tightly packed particles as compared to
structures with less compactness. The mass to radius ratio (also
known as compactness factor $(u(r))$ gauges the compactness of a
cosmic object. According to Buchdahl's study \cite{47}, the
compactness of a spherical structure obeys the limit
$u(r)<\frac{4}{9}$. Further, the compactness of an astrophysical
object influences the wavelength of nearby electromagnetic
radiations. The waves deviate from the straight path due to the
strong gravitational field of extremely dense structures. The
redshift in electromagnetic radiation is measured as
\begin{equation*}
Z(r)=\frac{1-\sqrt{1-2u}}{\sqrt{1-2u}},
\end{equation*}
which should not exceed the values 2 and 5.211 corresponding to
isotropic \cite{47} and anisotropic configurations \cite{48},
respectively. The plots in Figure \textbf{4} indicate that the
isotropic spherical system is more massive and compact as compared
to the anisotropic analog. Consequently, it exerts a stronger
gravitational field leading to higher redshifts. It is worthwhile to
mention here that the compactness and redshift parameters lie within
their respective admissible ranges.
\begin{figure}\center
\epsfig{file=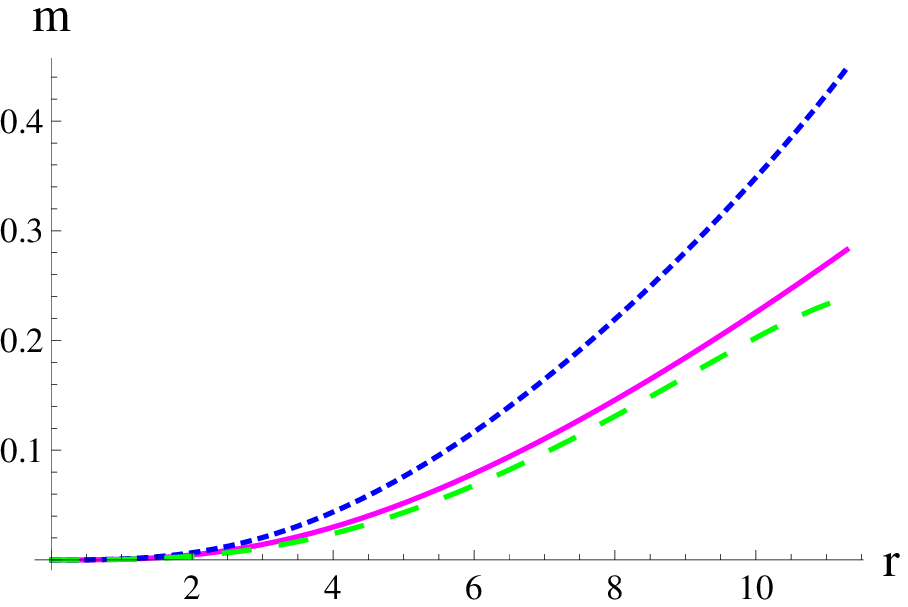,width=0.4\linewidth}\epsfig{file=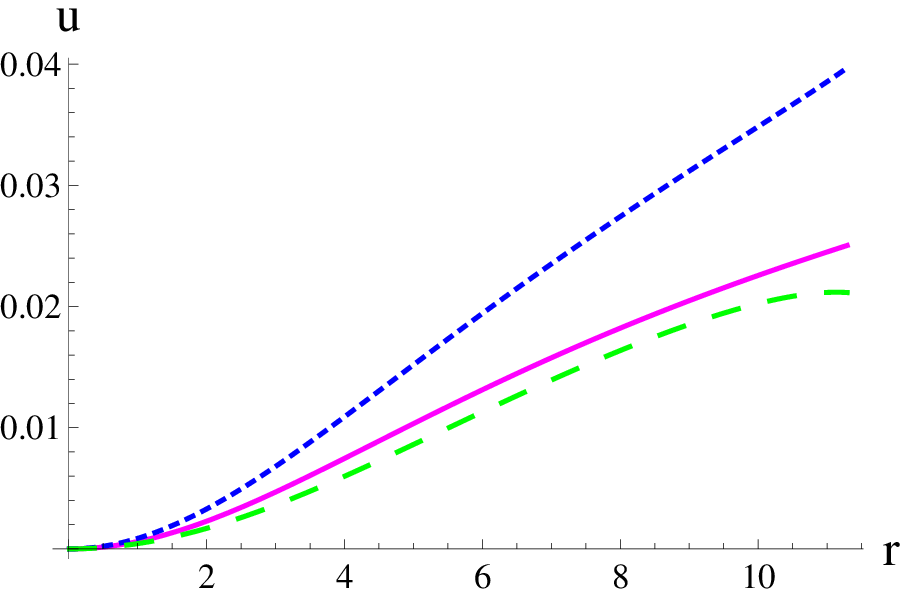,width=0.4\linewidth}
\epsfig{file=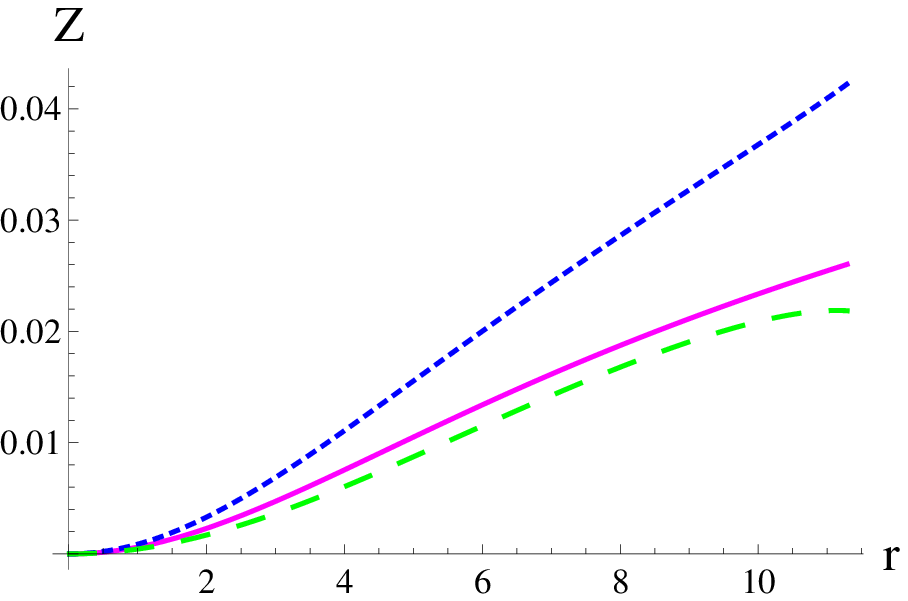,width=0.4\linewidth} \caption{Behavior of
$m,~u$ and $Z$ corresponding to solution I.}
\end{figure}
\begin{figure}\center
\epsfig{file=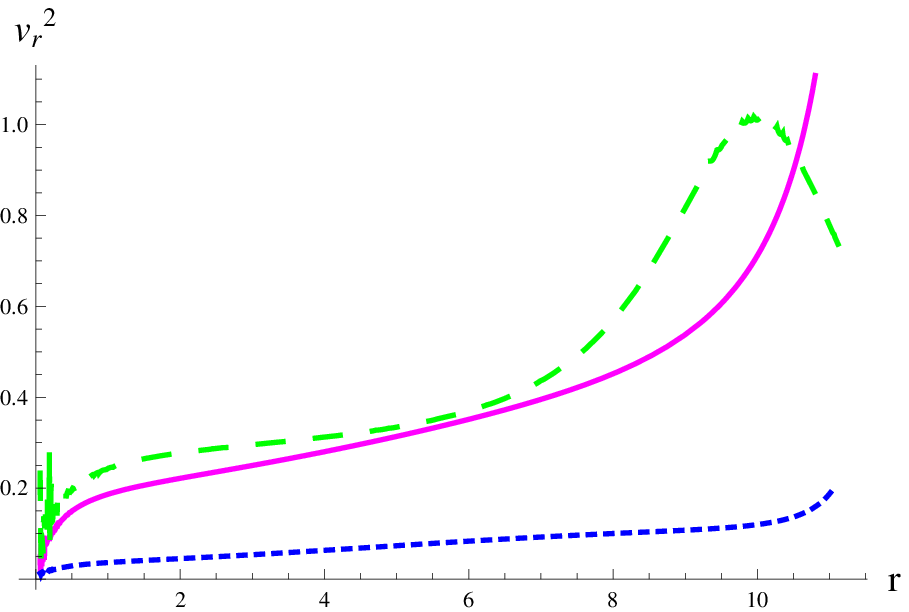,width=0.4\linewidth}\epsfig{file=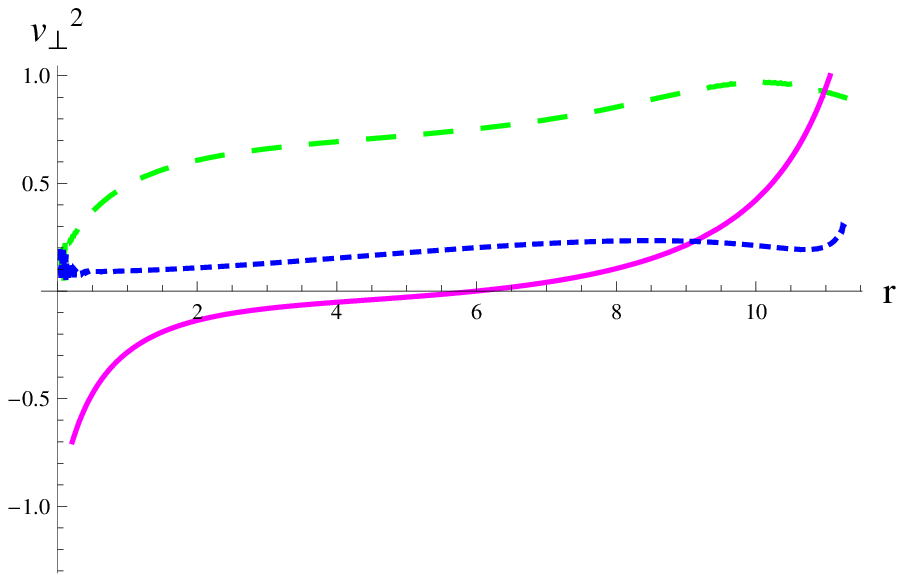,width=0.4\linewidth}
\epsfig{file=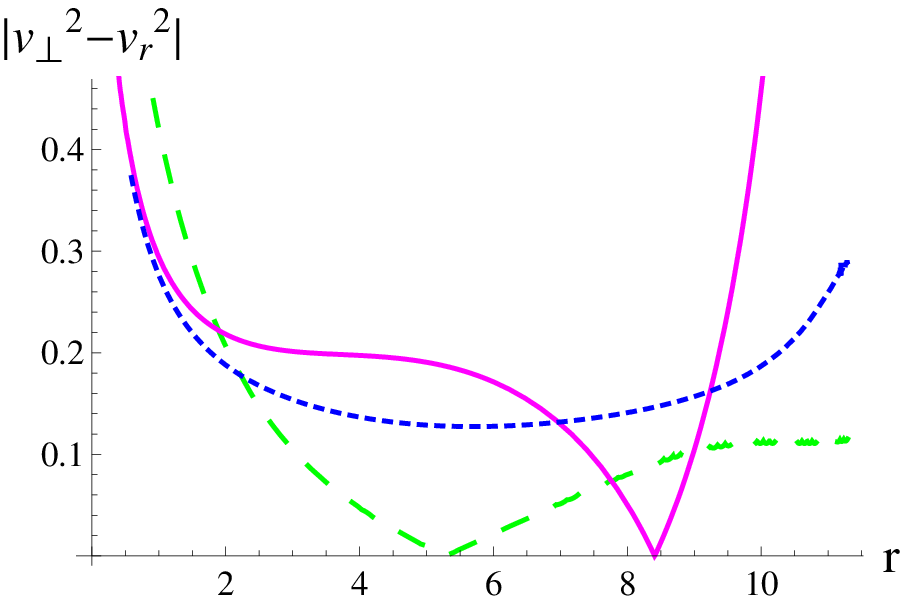,width=0.4\linewidth} \caption{Plots of
radial/tangential velocities and $|v_\perp^2-v_r^2|$ of solution I.}
\end{figure}

We check the stability of the current setup by calculating the
radial $(v_r^2=\frac{dp_r}{d\rho})$ and tangential
$(v_\perp^2=\frac{dp_\perp}{d\rho})$ speeds of sound traveling
through the considered medium. Sound wave must travel at a speed
less than that of light to maintain causality, i.e., $v_r^2$ and
$v_\perp^2$ must lie within the interval (0, 1). Figure \textbf{5}
illustrates that the celestial structure is unstable for $\eta=0.2$
as $v_\perp^2<0$. However, the system attains stability when $\eta$
increases. Herrera's cracking concept is an additional tool to
inspect the stability of the system. This criterion is fulfilled
when the components of sound speed satisfy the inequality
$0<|v_\perp^2-v_r^2|<1$. The static sphere is consistent with this
criterion for lower as well as higher values of $\eta$ as shown in
Figure \textbf{5}.

\subsection{Solution II}

In order to formulate the second solution, we specify
Eqs.(\ref{12})-(\ref{14}) via the metric potentials of Tolman IV
solution given as \cite{49}
\begin{eqnarray}\label{32}
\phi_1(r)&=&\ln\left(b^2
\left(\frac{r^2}{a^2}+1\right)\right),\\\label{33}
\lambda(r)&=&\left(\frac{\left(\frac{r^2}{a^2}+1\right)
\left(1-\frac{r^2}{f^2}\right)}{\frac{2 r^2}{a^2}+1}\right),
\end{eqnarray}
where the unknown constants $a,~b$ and $f$ are determined via a
singularity-free junction with Schwarzschild vacuum solution as
\begin{eqnarray}\label{34}
a^2&=&-\frac{R^2\left(M\left(28R-2
Rh+R^2h\right)-2M^2(12-\omega_{BD})-8
R^2\right)}{hR(R-2M)+2M^2\omega_{BD}},\\\nonumber
b^2&=&\frac{\left(3hR^2+6M^2(\omega_{BD}+12)-6M\left(R
h+10R\right)+8
R^2\right)(R-2M)}{2R\left(hR^2+2M^2(\omega_{BD}+12)-2M\left(R
h+11R\right)+4R^2\right)},\\\label{35}\\\nonumber
f^2&=&(m_\varrho^4R^6
+M^2\left(2h(\omega_{BD}+12+4m_\varrho^4R^4)+4R^2h+8(\omega_{BD}+12)\right)\\\label{36}
&-&4M\left(m_\varrho^4R^5+6R
h+16R\right))^{-1}(4R^3(3M-2R)\left(h+4\right)).
\end{eqnarray}
where $h=m_\varrho^2R^2\sqrt{1-\frac{2M}{R}}$. The field equations
related to the considered setup take the form
\begin{eqnarray*}
\rho&=&\left(2 f^2 r^2 \left(a^2+2 r^2\right)^2 \varrho
\right)^{-1}\left(r^2 \omega_{BD}  \left(a^2+2 r^2\right) \varrho
'^2 \left(\left(a^2+r^2\right)
\left(r^2-f^2\right)\right.\right.\\\nonumber &-&\left.\left.\eta
f^2 \left(a^2+2r^2\right) \mu(r)\right)+f^2 r^2 \left(a^2+2
r^2\right)^2 \varrho (r) V(\varrho )+2 \varrho^2 \left(r \left(r
\left(3 a^4\right.\right.\right.\right.\\\nonumber
&+&\left.\left.\left.\left.3 a^2f^2+7 a^2 r^2+2 f^2 r^2+6
r^4\right)-\eta f^2 \left(a^2+2 r^2\right)^2 \mu'(r)\right)-\eta f^2
\left(a^2\right.\right.\right.\\\nonumber &+&\left.\left.\left.2
r^2\right)^2 \mu(r)\right)-r \varrho (r) \left(2 r \left(a^2+2
r^2\right) \varrho''(r) \left(\eta  f^2 \left(a^2+2 r^2\right)
\mu(r)\right.\right.\right.\\\nonumber
&+&\left.\left.\left.\left(a^2+r^2\right)
\left(f^2-r^2\right)\right)+\varrho '(r) \left(\eta  f^2 r
\left(a^2+2 r^2\right)^2 \mu'(r)+4 \eta f^2
\left(a^2\right.\right.\right.\right.\\\nonumber
&+&\left.\left.\left.\left.2 r^2\right)^2 \mu+2 \left(a^4 \left(2
f^2-3 r^2\right)+a^2 \left(5 f^2 r^2-8 r^4\right)+4 f^2 r^4-6
r^6\right)\right)\right)\right),\\
p_r&=&-\frac{\varrho '(r) \left(\frac{\left(a^2+r^2\right)
\left(f^2-r^2\right)}{f^2 \left(a^2+2 r^2\right)}+\eta \mu(r)\right)
\left(r\omega_{BD}  \varrho '(r)-\frac{2 \left(2 a^2+3 r^2\right)
\varrho (r)}{a^2+r^2}\right)}{2 r \varrho (r)}\\\nonumber
&+&\frac{\varrho
(r)}{r^2}\left(\frac{\left(a^2+3r^2\right)\left(\frac{\left(a^2+r^2\right)
\left(f^2-r^2\right)}{f^2\left(a^2+2 r^2\right)}+\eta\mu(r)\right)}
{a^2+r^2}-1\right)-\frac{V(\varrho )}{2},\\
p_\perp&=&\left(2 f^2 r \left(a^2+r^2\right)^2 \left(a^2+2
r^2\right)^2 \varrho \right)^{-1}\left(r \omega_{BD}
\left(a^2+r^2\right)^2 \left(a^2+2 r^2\right) \varrho
'^2\right.\\\nonumber &\times&\left.\left(\eta  f^2 \left(a^2+2
r^2\right) \mu+\left(a^2+r^2\right)
\left(f^2-r^2\right)\right)-\left(a^2+2 r^2\right) \varrho^2
\left(\left(a^2+r^2\right)\right.\right.\\\nonumber
&\times&\left.\left.\left(2 r \left(a^2+r^2\right) \left(a^2-f^2+3
r^2\right)-\eta  f^2 \left(a^2+2 r^2\right)^2 \mu'(r)\right)-2 \eta
f^2 r \right.\right.\\\nonumber &\times&\left.\left.\left(2 a^4+5
a^2 r^2+2 r^4\right) \mu\right)+\left(a^2+r^2\right) \varrho
\left(\varrho ' \left(2 \eta  f^2 \left(a^2+2 r^2\right)^3
\mu(r)\right.\right.\right.\\\nonumber
&+&\left.\left.\left.\left(a^2+r^2\right) \left(\eta  f^2 r
\left(a^2+2 r^2\right)^2 \mu'(r)+2 \left(a^4 \left(f^2-2r^2\right)+3
a^2 r^2
\left(f^2\right.\right.\right.\right.\right.\right.\\\nonumber
&-&\left.\left.\left.\left.\left.\left. 2 r^2\right)+4 f^2 r^4-6
r^6\right)\right)\right)+2 r \left(a^4+3 a^2 r^2+2
r^4\right)\varrho''\left(\eta  f^2 \left(a^2+2
r^2\right)\right.\right.\right.\\\nonumber
&\times&\left.\left.\left.\mu(r)+\left(a^2+r^2\right)
\left(f^2-r^2\right)\right)\right)-f^2 r \left(a^4+3a^2 r^2+2
r^4\right)^2 \varrho (r) V(\varrho )\right),
\end{eqnarray*}

The closely linked features of intricate astrophysical objects lead
to complex behavior. Thus, it is difficult to determine the outcome
of a slight change in any of the state determinants. For this
purpose, it is convenient to formulate a complexity factor that
combines the essential properties of a spherical system in a single
equation. Recently, Herrera \cite{50} evaluated complexity of
anisotropic static as well as dynamical spheres via structure
scalars. These scalars were determined by splitting the Riemann
tensor orthogonally. This technique is based on the assumption that
the complexity reduces to zero if the fluid distribution is
isotropic and homogeneous in nature. The same scheme was adopted to
compute a complexity factor for static sphere in the presence of a
massive scalar field \cite{51}. This factor is defined in terms of
the following structure scalar
\begin{equation}\label{37}
Y_{TF}=Y_{TF}^{(m)}+Y_{TF}^{\varrho}=\frac{\Delta}{\varrho}
-\frac{1}{2r^3}\int_0^rr^3T^{0'\text{(eff)}}_{0}dr-
\frac{e^{-\phi_2}\varrho'}{2r\varrho},
\end{equation}
which effectively describes how Tolman mass $(m_T)$ corresponding to
the anisotropic sphere deviates from that of an isotropic system
(provided $\Delta\neq0$ and $\frac{d\rho}{dr}\neq0$) as
\begin{equation*}
m_{T}=(m_{T})_{\Sigma}\left(\frac{r}{r_{\Sigma}}\right)+r^3\int_{r}^{r_{\Sigma}}
\frac{e^{\frac{\phi_1+\phi_2}{2}}}{r}(-Y_{TF}^{(m)}+Y_{TF}^\varrho)+
\frac{e^{\frac{\phi_1-\phi_2}{2}}\varrho'}{2r\varrho}dr.
\end{equation*}
It must be noted that a complexity-free matter source may not be
isotropic and homogeneous. In the current scenario, decoupling the
complexity factor via Eq.(\ref{11}) leads to
\begin{eqnarray}\nonumber
Y_{TF}&=&\frac{p_\perp-p_r}{\varrho}
-\frac{1}{2r^3}\int_0^rr^3(\rho'+T^{0'\varrho}_{0})dr-
\frac{\lambda\varrho'}{2r\varrho}\\\label{38}
&+&\frac{\Theta_1^{1(\text{eff})}-\Theta_2^{2(\text{eff})}}{\varrho}-
\frac{1}{2r^3}\int_0^rr^3\Theta_0^{0'(\text{eff})}dr-\frac{\eta\mu\varrho'}{2r\varrho}.
\end{eqnarray}
The complexity factor can be split into two components
$\hat{Y}_{TF}$ and $Y_{TF}^\Theta$ which define the complexity of
original and additional sources, respectively as
\begin{eqnarray}\nonumber
\hat{Y}_{TF}&=&\frac{p_\perp-p_r}{\varrho}
-\frac{1}{2r^3}\int_0^rr^3(\rho'+T^{0'\varrho}_{0})dr-
\frac{\lambda\varrho'}{2r\varrho},\\\nonumber
{Y}_{TF}^\Theta&=&\frac{\Theta_1^{1(\text{eff})}-\Theta_2^{2(\text{eff})}}{\varrho}-
\frac{1}{2r^3}\int_0^rr^3\Theta_0^{0'(\text{eff})}dr-\frac{\eta\mu\varrho'}{2r\varrho}.
\end{eqnarray}
Thus, in the context of SBD gravity, the complexity factor of two
matter distributions existing within a static sphere is the sum of
their individual complexity factors.

In order to extend the solution corresponding to Tolman IV metric
functions, we impose the following constraint on complexity of the
additional source
\begin{equation}\label{42}
Y_{TF}^\Theta=0,
\end{equation}
which implies that the new source does not contribute to the
complexity of the seed distribution. The above condition,
corresponding to the current setup, turns out to be
\begin{eqnarray}\nonumber
&&\frac{\eta}{r \left(a^2+r^2\right) \varrho (r)}\left(r
\left(a^2+r^2\right) \left(r^3 \left(a^2+2
r^2\right)-\left(a^2+r^2\right) \varrho (r)\right)
\mu'(r)+\mu(r)\right.\\\label{43} &&\times\left. \left(r^3 \left(r
\left(a^2+r^2\right)^2 \varrho'-2\left(a^4+2 a^2 r^2+2
r^4\right)\right)-\left(a^2+r^2\right)^2 \varrho\right)\right)=0.
\end{eqnarray}
Numerical solutions of Eqs.(\ref{2*}) and (\ref{43}) for
$\eta=0.2,~0.5,~1$ with the initial conditions $\varrho(0)=0.25$ and
$\mu(0)=0$ determine the corresponding scalar field and deformation
function, respectively. Table \textbf{2} includes the values of
$\omega_{BD}$ and other important parameters. We graphically analyze
physical characteristics of solution II using the mass and radius of
PSR J1614-2230. Figure \textbf{6} displays the well-behaved metric
function after the transformation. The energy density and pressure
components, presented in Figure \textbf{7}, attain maximum values at
the center and remain positive throughout the stellar interior. The
anisotropy increases for some distance and then decreases towards
the boundary. Moreover, all state parameters vary directly with the
decoupling parameter. The considered system represents a viable
distribution as it complies with all the energy bounds (refer to
Figure \textbf{8}). Figure \textbf{9} shows that mass, compactness
and redshift increase with the radial coordinate. Further,
compactness and redshift parameters obey the corresponding upper
limits. Figure \textbf{10} indicates that the extended solution is
consistent with causality condition as well as cracking approach.
Thus, the stellar model corresponding to solution II is viable and
stable for chosen values of the decoupling parameter.
\begin{table} \caption{Values of physical parameters in solution II corresponding to
$\eta=0.2,~0.5,~1$.}
\begin{center}
\begin{tabular}{|c|c|c|c|}
\hline &$\eta=0.2$ & $\eta=0.5$ & $\eta=1$  \\
\hline $\omega_{BD}$ &9.87 & 10 & 10.2 \\
\hline $\rho_c~(gm/cm^3)$ & $9.398\times10^{15}$ &
$1.202\times10^{15}$ & $1.240\times10^{15}$\\
\hline$\rho_s~(gm/cm^3)$ & $2.008\times10^{15}$ & $2.6\times10^{15}$
&$2.6\times10^{15}$\\
\hline $p_c~(dyne/cm^2)$ & $1.754\times10^{36}$ &
$2.268\times10^{36}$ & $2.435\times10^{36}$\\
\hline$Z_s$ &0.057 & 0.072 & 0.072
\\\hline
\end{tabular}
\end{center}
\end{table}
\begin{figure}\center
\epsfig{file=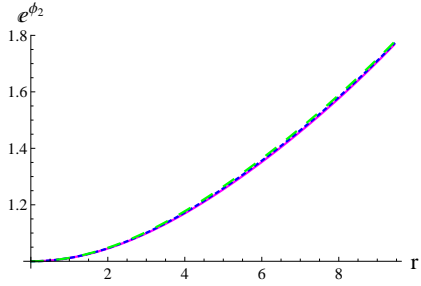,width=0.4\linewidth} \caption{Behavior of
deformed radial metric function for solution II.}
\end{figure}
\begin{figure}\center
\epsfig{file=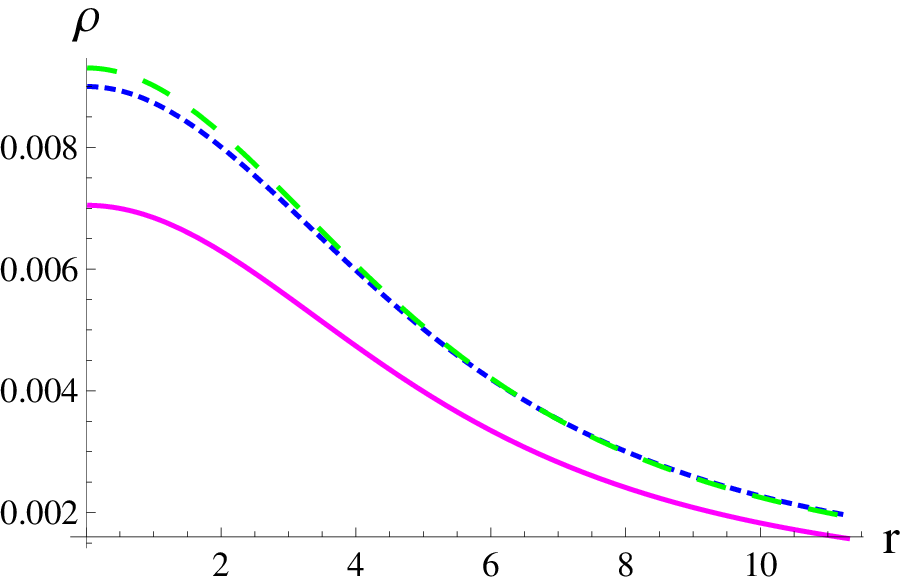,width=0.4\linewidth}\epsfig{file=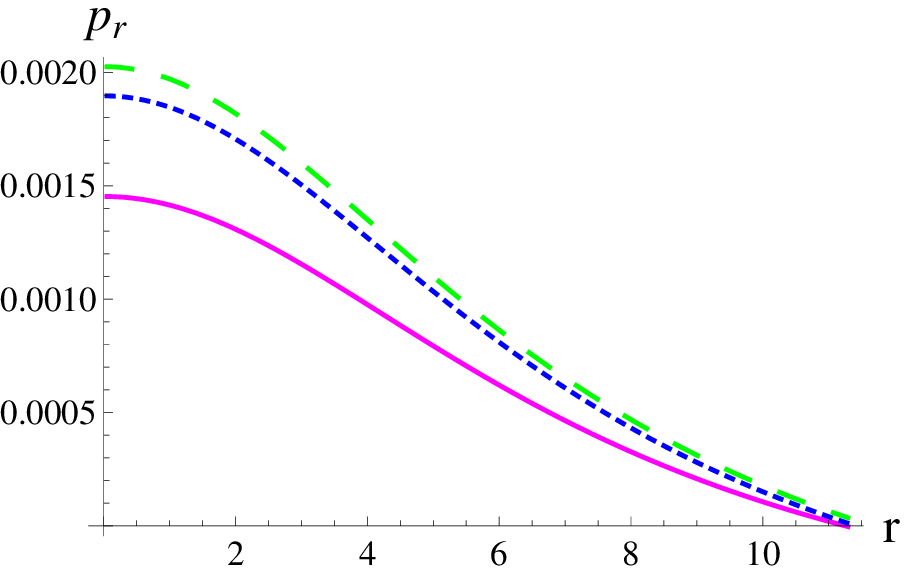,width=0.4\linewidth}
\epsfig{file=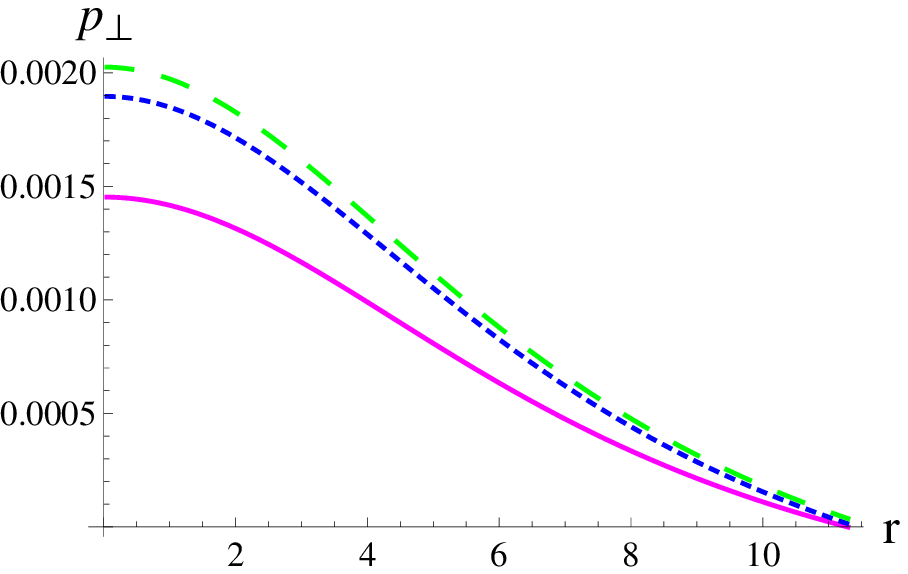,width=0.4\linewidth}\epsfig{file=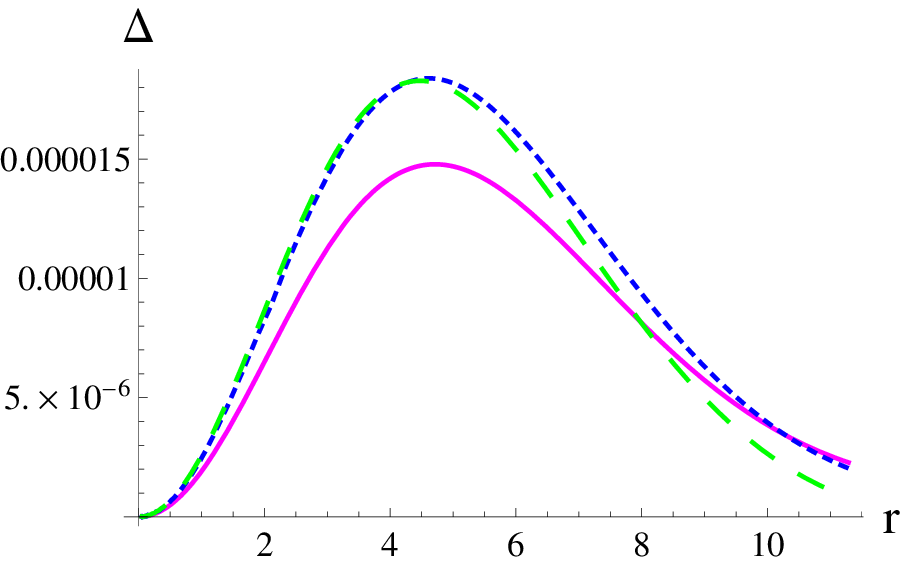,width=0.4\linewidth}
\caption{Behavior of $\rho,~p_r,~p_\perp$ (in $km^{-2}$) and
$\Delta$ for solution II.}
\end{figure}
\begin{figure}\center
\epsfig{file=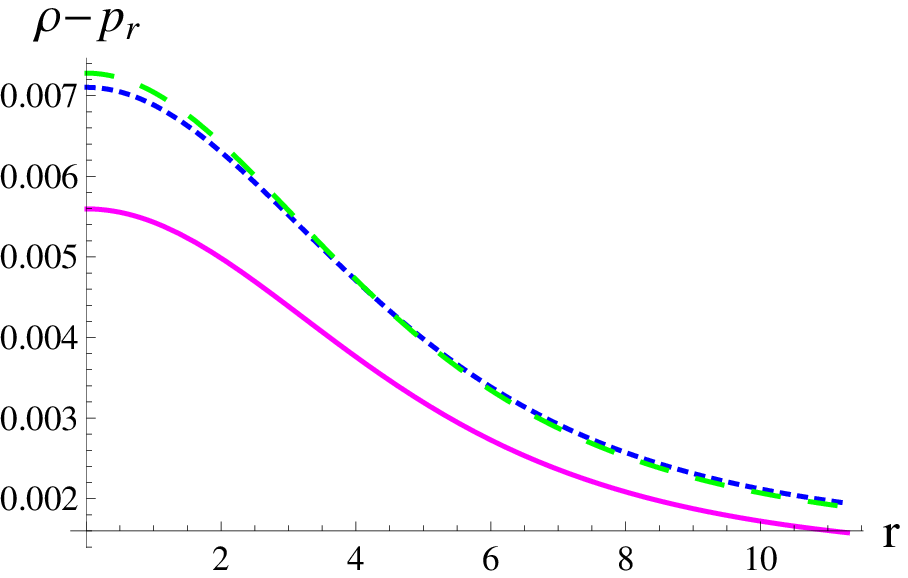,width=0.4\linewidth}\epsfig{file=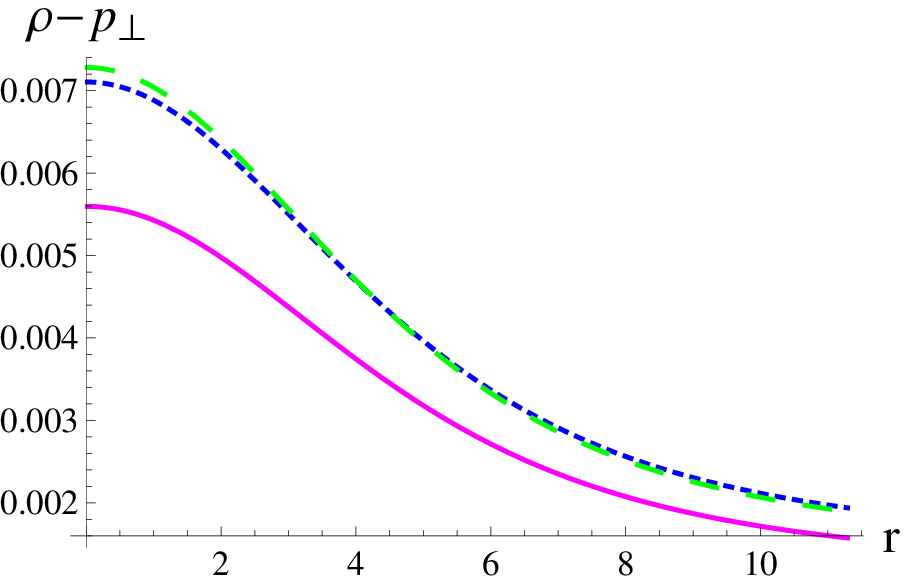,width=0.4\linewidth}
\caption{DEC for extended solution II.}
\end{figure}
\begin{figure}\center
\epsfig{file=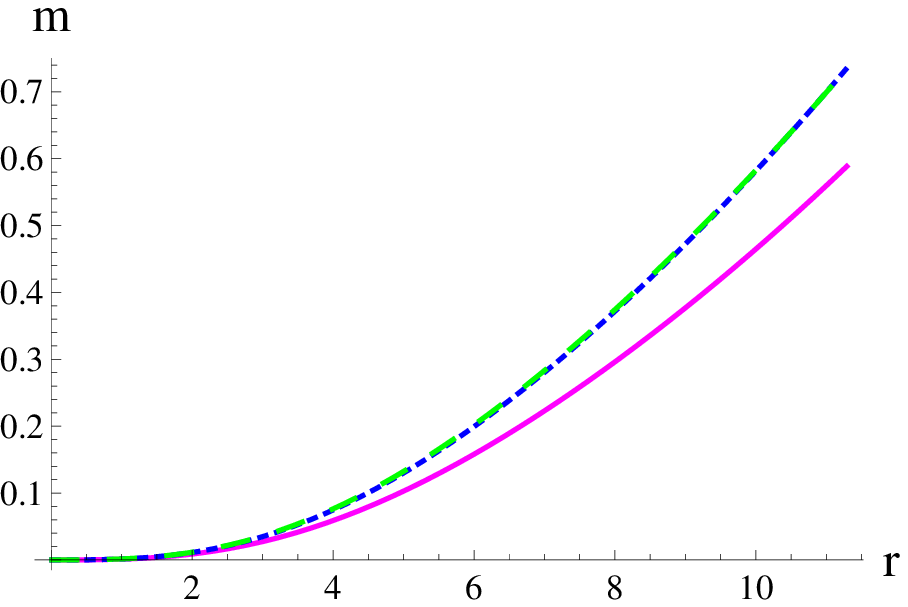,width=0.4\linewidth}\epsfig{file=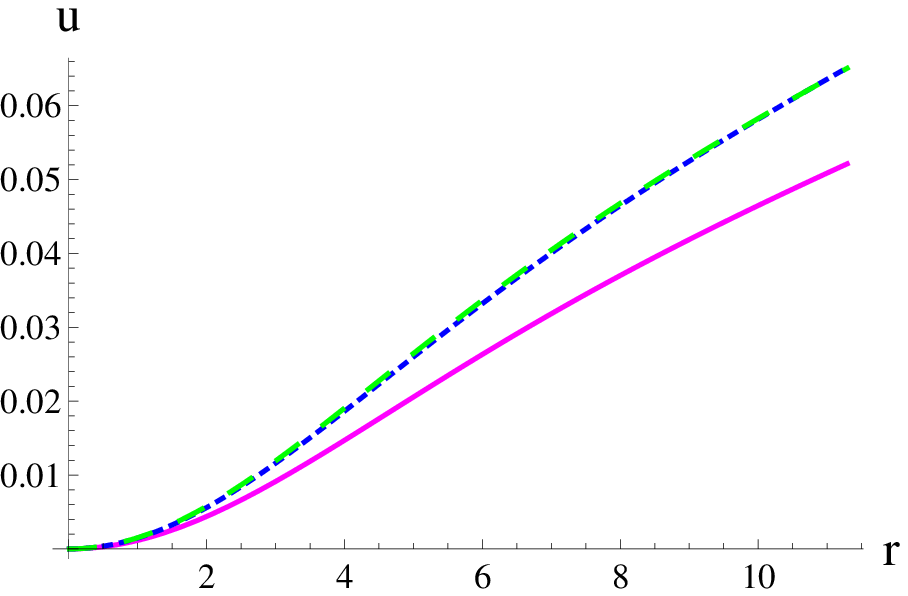,width=0.4\linewidth}
\epsfig{file=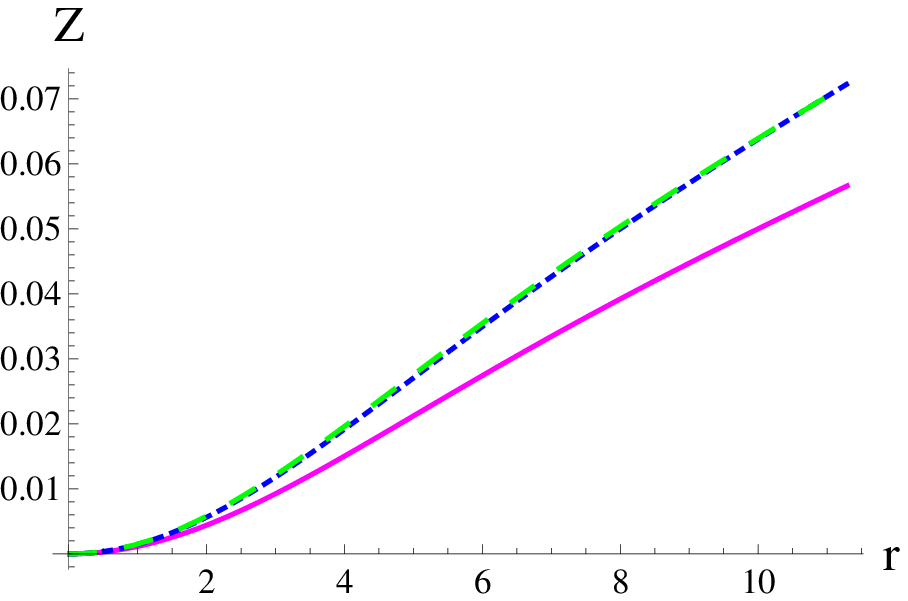,width=0.4\linewidth} \caption{Behavior of
$m,~u$ and $Z$ corresponding to solution II.}
\end{figure}
\begin{figure}\center
\epsfig{file=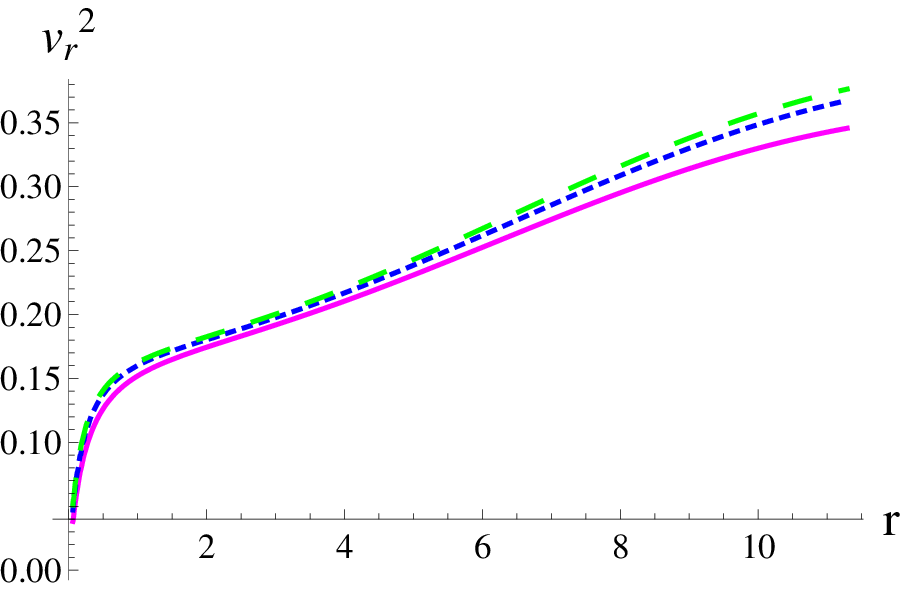,width=0.4\linewidth}\epsfig{file=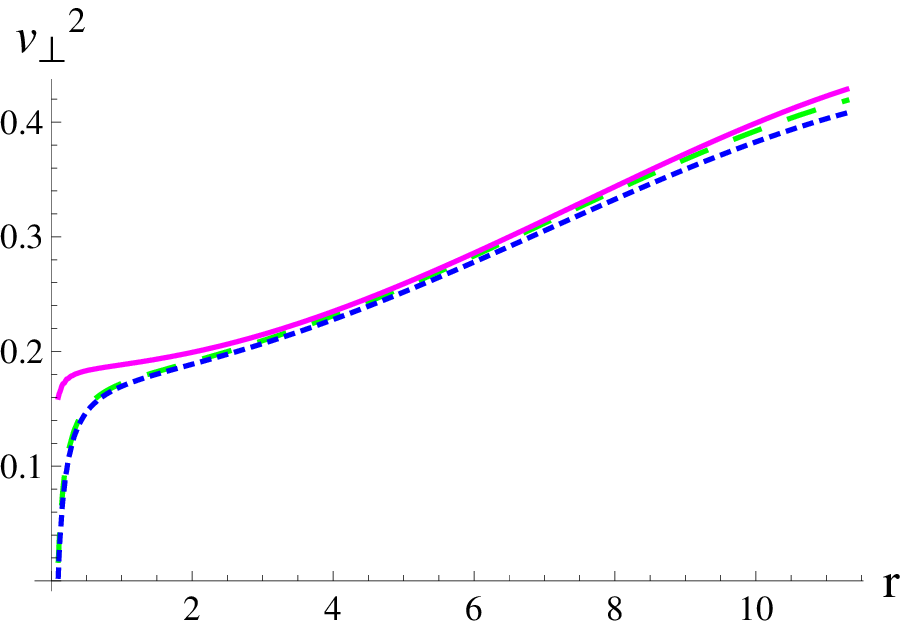,width=0.4\linewidth}
\epsfig{file=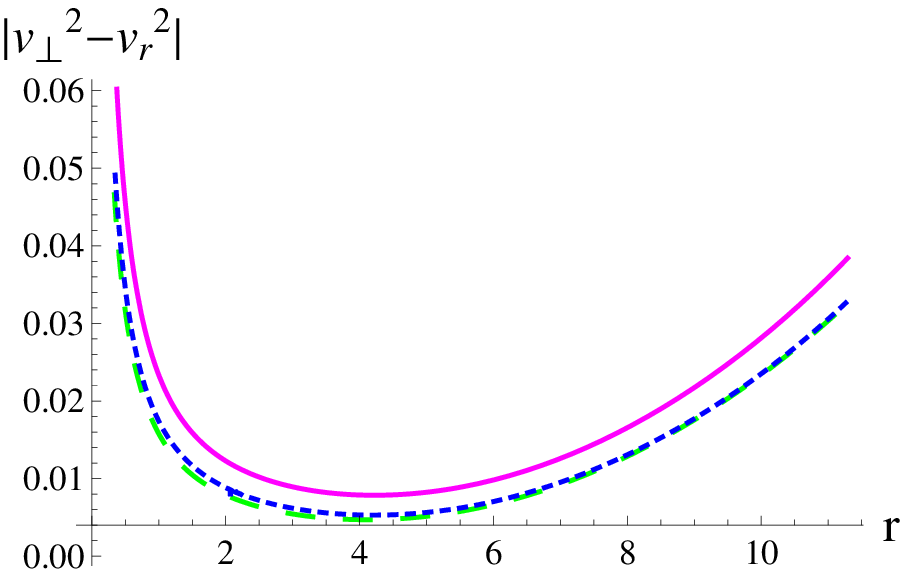,width=0.4\linewidth} \caption{Plots of
radial/tangential velocities and $|v_\perp^2-v_r^2|$ of extended
solution II.}
\end{figure}

\section{Discussion}

In this paper, we have generated new solutions representing
self-gravitating models in the framework of SBD gravity. For this
purpose, we have considered a static sphere in which two matter
distributions (anisotropic seed and additional sources) coexist.
Decoupling via MGD approach has been applied to disintegrate the SBD
field equations into two sets. The influence of the anisotropic
source is limited to the first array while the second system
exclusively incorporates the properties of the additional source. We
have used the ansatz $\phi_1(r)=\ln\left(B^2
\left(\frac{r^2}{A^2}+1\right)\right),~
\lambda(r)=\left(\frac{A^2+r^2}{A^2+3 r^2}\right)$ and metric
potentials of Tolman IV to specify the first system for solutions I
and II, respectively. Solution I has been obtained by applying the
condition $p_\perp-p_r=-(\Theta_1^1-\Theta_2^2)$ to construct
isotropic version of the original source. On the other hand, we have
constructed the second solution by assuming that the complexity of
the additional source is zero. These conditions have been solved
numerically alongside the wave equation to evaluate the deformation
function and scalar field. Moreover, the matching of internal and
external spacetimes has specified the unknown constants. Finally,
the behavior of obtained models has been checked via viability and
stability criteria for $V(\varrho)=\frac{1}{2}m_\varrho^2\varrho^2$
with $m_\varrho=0.01$.

We have analyzed the constructed solutions graphically for
$\eta=0.2,~0.5,~1$ by employing the mass and radius of PSR
J1614-2230. The geometric transformation in the radial metric
potential corresponding to solutions I and II has produced regular
stellar models free from any singularity. The pressure components
and energy density decrease away from the center for both solutions.
The matter variables related to solution II increase for higher
values of the decoupling parameter whereas the first stellar model
achieves maximum density and pressure for $\eta=0.5$ and $0.2$,
respectively. Moreover, we have observed that the anisotropy of the
first solution is negative for $\eta=0.2$. Thus, the stellar model
possesses outward repulsive force for $\eta=0.5$ only. On the other
hand, the anisotropy of solution II is positive for the considered
values of the parameters. However, it decreases towards the surface
indicating a weak repulsive force near the boundary. Furthermore,
both spherical structures are viable as they obey the energy
constraints.

The mass, compactness and redshift parameters attain maximum values
for $\eta=0.2$ and 1 in solutions I and II, respectively. However,
they do not cross their respective upper bounds in any
configuration. The isotropization of the anisotropic solution is
stable for $\eta=0.5,~1$ while the second self-gravitating model
agrees with the considered stability criteria (causality and
cracking approach) for all three values of the decoupling parameter.
Thus, it is concluded that the technique of decoupling can be
applied to obtain simplified versions of complex solutions in the
background of SBD theory. Moreover, the constraint $Y_{TF}^\Theta=0$
has yielded viable as well as stable stellar systems for a wider
range of the decoupling parameter as compared to the first solution.
It is interesting to mention here that the obtained results reduce
to GR for $\varrho=\text{constant}$ and $\omega_{BD}\rightarrow
\infty$.\\\\
\textbf{Acknowledgement} \\\\
One of us (AM) would like to thank the Higher Education Commission,
Islamabad, Pakistan for its financial support through the
\emph{Indigenous Ph.D. Fellowship, Phase-II, Batch-VI}.\\\\
\textbf{Data availability:} No new data were generated or analyzed
in support of this research.

\end{document}